\documentclass[12pt]{article}
\usepackage[a4paper, left=2cm,right=2cm]{geometry}
\usepackage{hyperref}
\usepackage{amsfonts} 
\usepackage{amsmath}
\usepackage{amssymb}
\usepackage{setspace}
\usepackage{color}

\usepackage{authblk}
\usepackage[utf8,applemac]{inputenc}
\usepackage{tensor}
\usepackage{cite}
\usepackage{tikz}
\usepackage{graphicx}% Include figure files
\usepackage{caption}
\usepackage{subcaption}

\usepackage{float}

\usepackage{bm}% bold math

\numberwithin{equation}{section}

\newcommand{\bea}{\begin{eqnarray}}
\newcommand{\eea}{\end{eqnarray}}
\newcommand{\be}{\begin{equation}}
\newcommand{\ee}{\end{equation}}

\renewcommand{\d}{\textrm{d}}

\begin{document}

\title{\Large\textbf{Quantum Corrected Geodesic Motion in Polymer Kerr-like Spacetime}}

\author{Zhiyang Guo\thanks{guozhiyang@s.ytu.edu.cn}}	
\author{Chen Lan\thanks{stlanchen@126.com}}
\author{Yan Liu \thanks{Corresponding author: yanliu@mail.bnu.edu.cn.}}

\affil{\normalsize{\em Department of Physics, Yantai University, 30 Qingquan Road, Yantai 264005, China}}

\date{} 

\maketitle

\begin{abstract}
\noindent

Rotating black holes are prevalent in astrophysical observations, and a Kerr-like solution that incorporates quantum gravity effects is essential for constructing realistic models. 
In this work, we analyze the geodesic motion of massive particles in a Kerr-like polymer spacetime, incorporating quantum corrections via a parameter $A_\lambda$. 
We demonstrate that increasing $A_\lambda$ allows for additional orbital evolution in extreme mass ratio inspiral (EMRI) systems before merging. 
Our results show that the radii, energy, and angular momentum of both the innermost stable circular orbit (ISCO) and marginal circular orbit (MCO) decrease as $A_\lambda$ increases. 
Furthermore, when the primary object becomes a wormhole, both prograde ISCO and MCO can intersect the transition surface at the wormhole throat and vanish as $A_\lambda$ grows. 
Additionally, we find that the eccentricity of periodic geodesic motion decreases monotonically with increasing $A_\lambda$. 
Finally, we explore the variation of the rational number that characterizes periodic motion and highlight the influence of the quantum parameter on different types of periodic orbits, classified by a set of integers associated with the rational number. 
This work contributes to the understanding of quantum gravity effects and offers potential observational signatures, particularly in the study of EMRIs.

\end{abstract}

%%%%%%%%%%%%%%%%%%%%%%%%%%%%%%%%%%%%%%%%%%%%%%%%%%%%%%%%%%%%%%%%%%%%%%%%%%%%%%%%%%%%%%%%

\newpage

\tableofcontents

%%%%%%%%%%%%%%%%%%%%%%%%%%%%%%%%%%%%%%%%%%%%%%%%%%%%%%%%%%%%%%%%%%%%%%%%%%%%%%%%%%%%%%%%
\section{Introduction}
%%%%%%%%%%%%%%%%%%%%%%%%%%%%%%%%%%%%%%%%%%%%%%%%%%%%%%%%%%%%%%%%%%%%%%%%%%%%%%%%%%%%%%%%

Black holes are one of the most intriguing predictions of general relativity. 
The first image of a black hole, captured by the Event Horizon Telescope (EHT) from the center of the M87 galaxy \cite{EventHorizonTelescope:2019dse}, provided direct evidence of their existence, making black holes a focal point of both theoretical physics and astrophysics. 
The strong gravitational field surrounding black holes gives rise to fascinating phenomena, such as gravitational lensing \cite{Soares:2024rhp,Soares:2023err,Bowman:2004ps}, black hole shadows \cite{EventHorizonTelescope:2019pgp, EventHorizonTelescope:2019ggy,Huang:2024wpj,Zhang:2023cuw,Su:2022roo,Li:2024abk,Guo:2018kis,Vagnozzi:2022moj,Afrin:2022ztr}, and gravitational waves produced by binary black hole mergers \cite{LIGOScientific:2016aoc}. 
Additionally, extreme mass-ratio inspirals (EMRIs), consisting of a stellar-mass black hole orbiting a supermassive black hole, are also a crucial source of gravitational waves. 
These systems are primary targets for future space-based gravitational wave detectors, such as LISA \cite{Danzmann:1997hm,LISA:2017pwj}, Taiji \cite{10.1093/nsr/nwx116}, and Tianqin \cite{TianQin:2015yph,Gong:2021gvw}. 
EMRIs emit low-frequency gravitational waves that encode information about the dynamics of the small-mass object and the spacetime of the supermassive black hole \cite{Jiang:2024lwg,Babak:2017tow}, providing insight into the nature of gravity and enabling further tests of general relativity  \cite{Fu:2024cfk,Yang:2024lmj,Yang:2024cnd}.

While general relativity successfully explains gravitational phenomena on macroscopic scales, it still faces unresolved challenges, such as the singularity problem \cite{Penrose:1964wq,Hawking:1970zqf,Lan:2023cvz}. 
General relativity breaks down when the spacetime curvature approaches the Planck scale. 
Therefore, integrating quantum mechanics, which excels at small scales, with general relativity is crucial to resolving this issue \cite{Adler:2010wf,Ng:2003jk}, and it has prompted the development of quantum gravity. 

Loop quantum gravity (LQG) is a leading candidate for developing a quantum theory of gravity \cite{Rovelli:1997yv}. 
It is a non-perturbative approach that addresses classical singularities in both cosmology and black hole spacetimes. 
Over the past few decades, significant progress has been made in addressing cosmological singularities, and extensive research has been conducted on LQG theory and its applications \cite{Bojowald:2001xe,Bojowald:2005epg,Ashtekar:2006uz}. 
The LQG effective equations for static, spherically symmetric, non-rotating spacetimes have been well-established, leading to a comprehensive LQG solution for Schwarzschild-like black holes. 
However, challenges remain in solving LQG equations for axisymmetric rotating spacetimes, particularly when using real-valued Ashtekar-Barbero variables, leaving the spacetime associated with Kerr black holes largely unexplored. 
However, from a phenomenological perspective, rotating Kerr black holes are more relevant, as most observed black holes possess non-zero angular momentum \cite{Lan:2024wfo}. 
In 2020, Suddhasattwa et al. extended the polymer black hole model in LQG to rotating black holes using the Newman-Janis method \cite{Newman:1965tw} and proposed a framework for testing LQG in astrophysical settings \cite{Brahma:2020eos}. 
Following this, dynamic studies were conducted \cite{Xia:2023zlf}, exploring phenomena such as the weak gravitational deflection of massive particles \cite{Huang:2022iwl}.

Following the detection of gravitational waves and the future missions of space-based gravitational wave detectors, timelike geodesics have gained significant attention as a tool for exploring the sources of these waves. 
Periodic orbits are particularly valuable because they encode fundamental information about orbits around black holes. 
A set of integers $(z, w, v)$ was proposed \cite{Levin:2008mq} to define a rational number $q$, which serves as an index for periodic orbits around black holes. 
In this context, $z$ represents the number of leaves in a complete periodic orbit, $w$ indicates the number of additional rotations that the particle makes around the central black hole within a radial cycle, and $v$ indicates the next apocenter that the particle will attain after departing from its initial apocenter position (denoted as $v=0$). 
This framework connects rational numbers and periodic orbits based on the topological features of the orbit and the frequencies of its radial and angular motions. 
This taxonomy significantly facilitates the study of periodic orbits and has been applied to various black holes  \cite{Misra:2010pu,Liu:2018vea,Shabbir:2025kqh,Tu:2023xab,Babar:2017gsg,Wang:2022tfo,Zi:2024jla,Lim:2024mkb,Junior:2024tmi,Meng:2024cnq,Zhao:2024exh,Jiang:2024cpe,Yang:2024lmj,QiQi:2024dwc,Uktamov:2024zmj,Li:2024tld,Lin:2023rmo,Lin:2022llz,Lin:2022wda,Lu:2025cxx,Chen:2025aqh}.
Besides, studying the geodesic motion of massive particles around black holes allows for a deeper exploration of their spacetime structure \cite{Liu:2024pui,Huang:2023yqd,Zhang:2025lhm,Battista:2022krl}. 
In this work, we investigate the timelike geodesic motion in the polymer Kerr-like spacetime given by \cite{Brahma:2020eos}.

The structure of the paper is as follows: In Sec.\ \ref{sec2}, we review the solution for rotating polymer black holes in LQG. 
In Sec.\ \ref{sec3}, we analyze geodesic motion around black hole spacetimes, derive the radial potential for massive test particles, and investigate circular orbits, MCOs, and ISCOs. 
By comparing the characteristics of prograde and retrograde orbits, we reveal interesting properties of quantum corrections. In Sec.\ \ref{sec4}, we explore the impact of quantum corrections, energy, and angular momentum on the eccentricity and rational numbers of periodic orbits, illustrating the influence of each parameter. 
Finally, we present our conclusions and discussion in Sec.\ \ref{sec5}. 
In App.\ \ref{appendix:A}, we detail the calculations for the MCO orbits, while in App.\ \ref{appendix:B} we discuss the rational numbers used for retrograde periodic orbits. 
Throughout the paper, we use the geometrized unit system with $G = c = 1$ and adopt the metric convention $(- + + +)$.

%%%%%%%%%%%%%%%%%%%%%%%%%%%%%%%%%%%%%%%%%%%%%%%%%%%%%%%%%%%%%%%%%%%%%%%%%%%%%%%%%%%%%%%%
\section{Overview of Rotating Black Holes in Polymer Loop Quantum Gravity}
\label{sec2}
%%%%%%%%%%%%%%%%%%%%%%%%%%%%%%%%%%%%%%%%%%%%%%%%%%%%%%%%%%%%%%%%%%%%%%%%%%%%%%%%%%%%%%%%

\begin{figure}[!ht]
    \centering
    \includegraphics[width=0.45\linewidth]{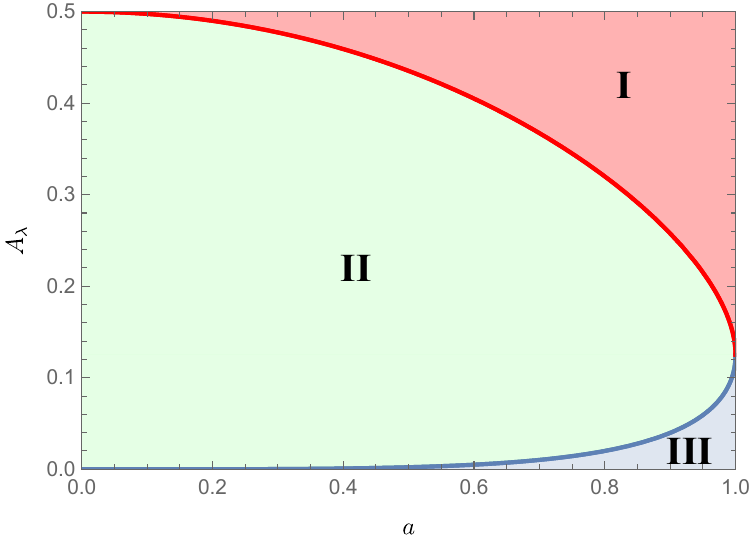}
    \captionsetup{width=.9\textwidth}
    \caption{The spacetime structure of the black hole determined by the parameter space $(a, A_\lambda)$, the red line represents the transition surface located on the outer horizon, the blue line represents the transition surface located on the inner horizon.}
    \label{fig:horizon}
\end{figure}

In this section, we review rotating black holes within the framework of polymer loop quantum gravity. 
The rotating black hole model discussed here was first proposed by \cite{Brahma:2020eos},
\begin{equation}
\d s^2 = -  \left( 1-\frac{2 M b}{\rho^2}  \right) \d t^2
	-\frac{4aMb\sin^2{\theta}}{\rho^2}\d t \d\varphi
		+\rho^2d\theta^2+\frac{\rho^2 dr^2}{\Delta}
			+\frac{\varSigma \sin^2{\theta}}{\rho^2}\d\varphi^2,
			\label{sol}
\end{equation}
where 
\bea
&&\rho^2 = b^2 + a^2 \cos^2{\theta}, \qquad 
	M = b\left(1 - 8A_\lambda M^{2}_{b} \tilde{a}\right)/2 , \\
&&\Delta = 8A_\lambda M^{2}_{b} \tilde{a} b^2 + a^2 ,\qquad  
	\Sigma = (b^2 + a^2)^2 - a^2 \Delta \sin^2{\theta},
\eea
and the terms $ \tilde{a} $ and $ b $ are defined as
\begin{eqnarray}
 b(r)^2 &=& \frac{512A_{\lambda}^{3}M_{b}^{4}M_{w}^{2}
 	+\left(r+\sqrt{8 A_\lambda M_{b}^{2}+r^2}\right)^6}{8 \sqrt{8 A_\lambda M_{b}^{2}+r^2}\left(r+\sqrt{8 A_\lambda M_{b}^{2}+r^2}\right)^3}, \\
\tilde{a} &=& \frac{1}{b^2(r)}\left(1+\frac{r^2}{8 A_\lambda M_{b}^{2}}\right)
	\left(1-\frac{2M_b}{\sqrt{8 A_\lambda M_{b}^{2}+r^2}}\right),
\end{eqnarray}
where the parameters $a$ and $b$ share the same dimension of mass, while $\tilde a$ carries the dimension of mass$^{-2}$.
 The parameter $ A_\lambda $ is defined as $ A_\lambda \equiv \frac{1}{2}(\lambda_k/(M_b/M_w))^{\frac{2}{3}}  $, where $ \lambda_k $ represents a quantum correction parameter. 
Consequently, $A_\lambda$ and $\lambda_k$ both are dimensionless.
 The position of the transition surface is defined at $ r = 0 $, which represents an interface that smoothly connects an axisymmetric black hole to a white hole, characterized by the respective masses $ M_b $ and $ M_w $. For simplicity, we assume that the black hole and white hole have equal masses, $ M_b = M_w $,  and render all variables dimensionless by setting mass to unity, $ M_b = M_w=1$.
 %normalizing them with respect to $M_b$}. 
 By solving $ \Delta = 0 $, the inner and outer event horizons of the rotating black hole, denoted by $ r_- $ and $r_+$ can be obtained, which are given by 
\begin{equation}
 r_{\pm}^{2}=\left( 1 \pm \sqrt{ 1 - a^2 } \right)^2 - 8 A_\lambda ,
    \label{eq:(2.4)}
\end{equation}
 where all quantities are dimensionless under our chosen normalization. As $a \to 0$, the inner horizon vanishes, and the outer horizon reduces to $ r_+|_{a \to 0} = 2 \sqrt{1 - 2A_\lambda} $, 
corresponding to the case of a non-rotating polymer black hole in Ref.\ \cite{Tu:2023xab}. 
By illustrating the inner and outer horizons in the parameter space of $ (a, A_\lambda)$ with $\theta=\pi/2$, 
the three types of the spacetime structure of a polymer Kerr-like black hole can be obtained as shown in Fig.\ \ref{fig:horizon}. 

In region I, the Eq.\ \eqref{sol} describes a rotating timelike wormhole without an event horizon implicated by the imaginary $r_\pm$, 
where the transition surface is located at $r=0$ in the wormhole throat. 
In region II, the black hole has a single outer horizon, and the transition surface becomes spacelike, 
located inside the outer horizon. In region III, the black hole possesses both an inner and an outer horizon, 
with its classical ring singularity replaced by a timelike transition surface located inside the inner horizon. 
Note that on the red and blue curves, the transition surface is located explicitly on the outer and inner horizons, respectively. 
Subsequently, we specify the location of the ergosphere given by
\begin{equation}
    r^{\pm}_{s} = \sqrt{   \left(1 \pm \sqrt{1 - a^2 \cos^2{\theta}  }  \right)^2  - 8 A_\lambda },
\end{equation}
where the $ + $ and $ - $ signs respectively represent the inner and outer ergosphere. Then the ergoregion is defined by the radial ranges $ r_+ < r < r_s^+ $ and $ r_s^- < r < r_- $.

%%%%%%%%%%%%%%%%%%%%%%%%%%%%%%%%%%%%%%%%%%%%%%%%%%%%%%%%%%%%%%%%%%%%%%%%%%%%%%%%%%%%%%%%
\section{Circular Geodesics in Polymer-Corrected Kerr-like Spacetimes}
\label{sec3}
%%%%%%%%%%%%%%%%%%%%%%%%%%%%%%%%%%%%%%%%%%%%%%%%%%%%%%%%%%%%%%%%%%%%%%%%%%%%%%%%%%%%%%%%

In the following of this paper, we consider equatorial motion with $ \theta = \pi/2 $, 
and replace $ r $ with $ x = \sqrt{r^2 + 8A_\lambda} $ for simplicity. 
The dual vector of the 4-velocity is given by $ u_\mu = (-E, u_x, 0, L) $, 
where $E$ and $L$ represent the reduced energy and angular momentum of the orbital motion, respectively. 
Using the normalization condition $ u^\mu u_\mu = -1 $, we obtain
\bea
u_x=\pm\frac{x\sqrt{R(x)}}{r \Delta(x)},
\eea
where the $+$ and $-$ signs denote the outgoing and ingoing trajectories, $\Delta(x)=x^2-2x+a^2$ with the replacement, 
and the radial potential $R(x)$ is expressed as a fourth order polynomial given by
\bea
R(x)&=&(E^2-1)x^4+2x^3+[(E^2-1)(a^2-6A_\lambda) - L^2 -6 E^2 A_\lambda]x^2\nonumber\\
&+&2(a^2 E^2 - 2 a E L + L^2 - 6 A_\lambda)x+
	6 \left[a^2 (1 - 2 E^2) + 2 a E L + 6 E^2 A_\lambda\right] A_\lambda.
\label{eq:(3.2)}
\eea

Note that in the new coordinate system $(t, x, \theta, \phi)$, the inner and outer event horizons locate at $x_\pm=1\pm\sqrt{1-a^2}$, 
the ergosphere locates at $x_{ergo^\pm}=1\pm \sqrt{1-a^2 \cos^2\theta}$, 
and the transition surface $r=0$ turns to locate at  $x=\sqrt{8A_\lambda}$. 
The four velocity are instead given by
\bea
u^t&=&\frac{a^2 E [x (2 + x) - 12 A_\lambda] + 
 E (x^2 - 6 A_\lambda)^2 + 
 a L (-2 x + 6 A_\lambda)}{(x^2 - 
   6 A_\lambda)\Delta},\\
u^x&=&\pm\frac{\sqrt{R(x)}}{x^3 - 6 x A_\lambda},\\
u^\theta &=&0,\\
u^\phi &=&\frac{L (-2 + x) x + a E (2 x - 6 A_\lambda)}{(x^2 - 
   6 A_\lambda)\Delta}.
\eea

\begin{figure}[h!t]
    \centering
    \includegraphics[width=0.5\linewidth]{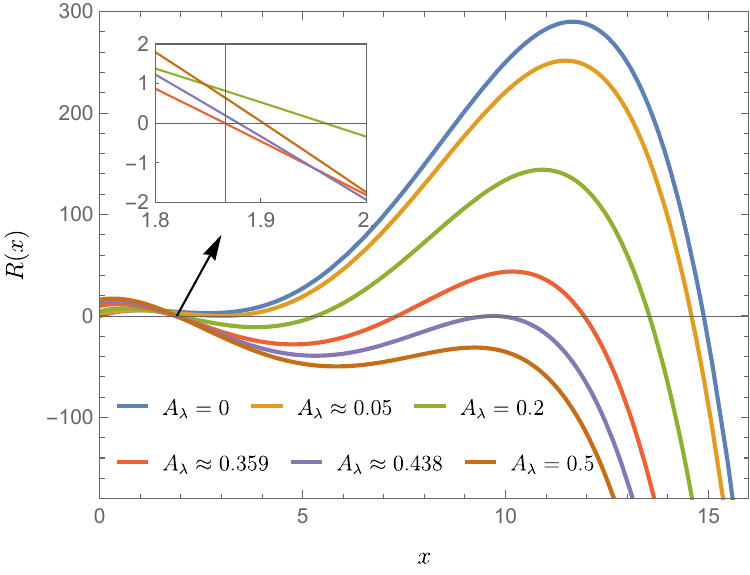}
    \captionsetup{width=.9\textwidth}
    \caption{The radial potential for a massive test particle around a rotating LQG polymer black hole for different values of $A_\lambda$. 
    Parameters are fixed at $E = 0.95$, $a = 0.5$, and $L = 3$.}
    \label{fig:RadialA}
\end{figure}

In Fig.\ \ref{fig:RadialA}, we demonstrate the structure of the radial potential with different quantum parameter $A_\lambda$, 
by fixing $ a = 0.5 $, $ E = 0.95 $, and $ L = 3 $. 
The blue curve with $A_\lambda=0$ denotes the radial potential of the classical Kerr case, which has only one real root outside of the horizon, corresponding to the turning point of the trapped orbit. As $A_\lambda$ increases, the double root structure emerges, such as the orange curve with $A_\lambda \approx 0.05$ exhibiting the unstable circular motion, and the purple curve with $A_\lambda \approx 0.438$ exhibiting the stable circular motion. The green curve with $A_\lambda=0.2$ corresponds to the root structure with three separated roots, exhibiting a generic bound motion and a trapped orbital motion. The red curve with $A_\lambda \approx 0.359$ exhibits a bound motion as well, but its first single root is located at the outer event horizon.

In the inset of Fig.\ \ref{fig:RadialA}, we show that as $A_\lambda$ increases, the turning point of the trapped orbit moves closer to the horizon at first, and then moves away after touching the horizon. 
However, in the ergoregion, the angular momentum $L$ of the orbital motion should obey \cite{Compere:2021bkk} 
\bea
L\leq L_+=\frac{2E}{a}\left(1 + \sqrt{1 - a^2} - 3 A_\lambda\right),
\label{L+}
\eea
where $L_+$ is the angular momentum of the horizon sliding orbital motion, as described by the red curve case in Fig.\ \ref{fig:RadialA}. 
Therefore, with the quantum correction, the trapped orbits could be disallowed even though the radial potential is positive, as in the case of the brown curve with $A_\lambda=0.5$ in the inset of Fig.\ \ref{fig:RadialA}. 
The constraint in Eq.~\eqref{L+} narrows the allowed range of orbital angular momentum compared to the classical Kerr case, implying that orbital trajectories that would plunge into a Kerr black hole may be prevented from doing so due to quantum corrections.
As a result, the inspiraling system must complete more orbits and radiate additional angular momentum to satisfy the constraint in Eq.~\eqref{L+} before transitioning to the plunge phase.

The circular orbital motion should satisfy the conditions $R(x) = 0$ and $R'(x) = 0$. 
Such that the angular momentum and energy of circular orbits are given by  
\bea
L_{cir\pm} &=& \frac{(6 a A_\lambda  - 2 a x) E_{cir\pm}  \pm \sqrt{\Delta( x^2 - 6 A_\lambda)(E_{cir\pm}^{2} (x^2 -6 A_\lambda)- x^2 +2 x   )}}{( x- 2) x},\label{eq:Lcir}
\\
 E^{2}_{cir\pm} &=& \frac{[a \sqrt{H} \pm ( x - 2 ) x^{3/2}]^2 [  x^2( x - 2 ) \mp 2 a \sqrt{H x} - H ] }{( x^2 - 6 A_\lambda )[  x^4( x - 2 )^2 +  (x^2( 5 - 2 x )- 6 A_\lambda ( x - 1 )  - 4 a^2 x)H   ] },\label{eq:Ecir}\\
H &=& 6 A_\lambda - 6 A_\lambda x + x^2,
\eea
where the $ + $ and $ - $ signs respectively represent the prograde and retrograde orbital motion.

\begin{figure}[!htb]
     \centering
     \begin{subfigure}[b]{0.45\textwidth}
         \centering
         \includegraphics[width=\textwidth]{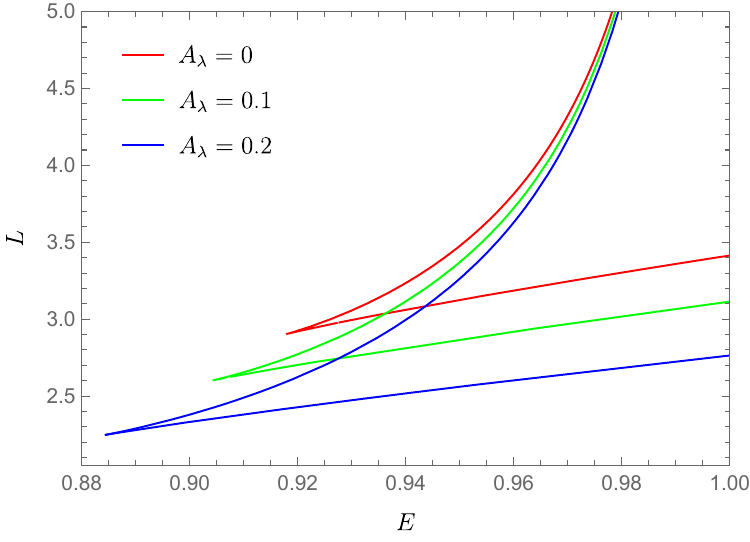}
         \caption{}
         \label{fig:allow}
     \end{subfigure}
     %\hfill
     \begin{subfigure}[b]{0.45\textwidth}
         \centering
         \includegraphics[width=\textwidth]{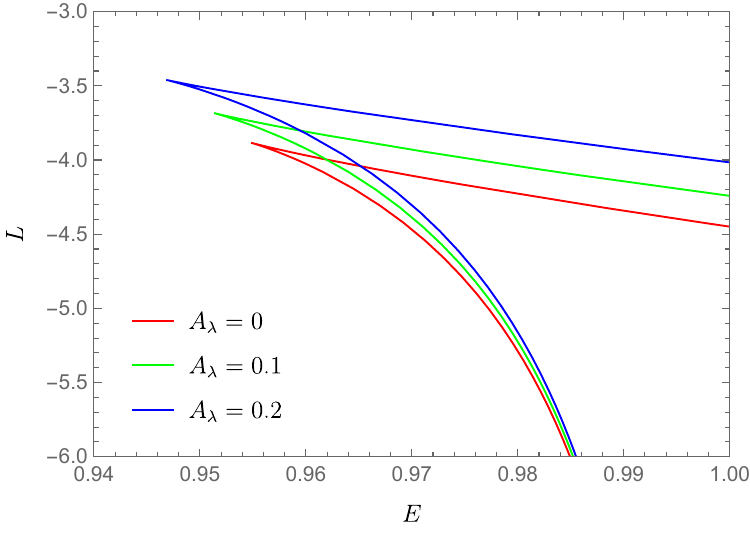}
         \caption{}
         \label{fig:allow-n}
     \end{subfigure}
      \captionsetup{width=.9\textwidth}
       \caption{Allowed parameter space for orbital angular momentum and energy of bound orbits for different values of $A_\lambda$. (a): the prograde motion, (b): the retrograde motion. }
        \label{fig:allowall}
\end{figure}

In Fig.\ \ref{fig:allowall}, we present the prograde and retrograde circular motion around the LQG black hole in the phase space of $(E, L)$. 
Each curve denotes the circular motion with specific quantum parameter $A_\lambda$, 
where the nearly straight part describes the unstable circular orbits and the left curved part describes stable circular orbits, 
separated by the vertex denoting the ISCO. 
The bound orbits are constrained in the region surrounded by the circular motion and the marginal orbits with $E=1$. 
It shows that with fixed orbital energy, the bound orbital motion has a lower angular momentum boundary as the parameter $A_\lambda$ increases. 
By comparing the spacing among these curves, we find that the unstable circular motion is much more sensitive to quantum correction than the stable circular motion, and the prograde orbital motion is more sensitive than the retrograde motion. 
That is because the quantum effect is much more obvious in the strong gravitational field near the black hole than in the weak field farther away.

%%%%%%%%%%%%%%%%%%%%%%%%%%%%%%%%%%%%%%%%%%%%%%%%%%%%%%%%%%%%%%%%%%
\subsection{Marginally Circular Orbits and Their Quantum Corrections}
 %%%%%%%%%%%%%%%%%%%%%%%%%%%%%%%%%%%%%%%%%%%%%%%%%%%%%%%%%%%%%%%%%%
 
For marginal orbits with $E=1$, the radial potential is reduced to a third order polynomial given by
\bea
R_m(x)=L^2 (2- x) x  -4 a L (x-3 A_\lambda)+
 2( x - 3 A_\lambda) (x^2 +a^2- 
    6 A_\lambda).
\eea
Considering the circular motion with $R_m(x)=0$ and $R_m'(x)=0$, 
the location and angular momentum of the marginal circular orbits (MCO) satisfy
\bea
0&=&4 a^2 (6 A_\lambda ( x_{m\pm} - 1) - x_{m\pm}^2)^2 (x_{m\pm} (a^2 + 3 ( x_{m\pm} - 1) x_{m\pm}) - 6 A_\lambda ( x_{m\pm}^2 - 1 )) \nonumber\\
    &+& (a^2 (6 A_\lambda + x_{m\pm}^2 - 
      6 A_\lambda x_{m\pm}^2 + x_{m\pm}^3) - ( x_{m\pm} - 1) (-36 A_\lambda^2 ( x_{m\pm} - 1)\nonumber\\ 
      &+& 12 A_\lambda x_{m\pm}^2 + ( x_{m\pm} - 4 ) x_{m\pm}^3))^2,\\
L_{m\pm} &=&\frac{6 a A_\lambda - 2 a x_{m\pm} \pm
 \sqrt{2 \Delta(x_{m\pm}) (18 A_\lambda^2 + x_{m\pm}^3 - 
     3 A_\lambda x_{m\pm} (2 + x_{m\pm}))}}{( x_{m\pm} - 2) x_{m\pm}},
\eea
where the $+$ and $-$ signs in the lower index denote prograde orbits and retrograde orbits separately. 
By treating $A_\lambda$ as a small quantity, the expressions for $ \Delta(x_{m\pm}) $, 
as well as the solutions for $ x_{m\pm} $ and $ L_{m\pm} $, are given by
\bea
\Delta(x_{m\pm}) &=& a^2 + ( x_{m\pm} - 2) x_{m\pm}\\
x_{m\pm}&=&2 + 2  \sqrt{1 \mp a} \mp a -3A_{x\pm}(a) A_\lambda+ O(A_\lambda^2),\label{eq:3.14}\\
L_{m\pm}&=&\pm2(1\pm\sqrt{1\mp a})+3A_{L\pm}(a) A_\lambda+O(A_\lambda^2),\label{eq:3.15}
\eea
where the coefficients $A_{x\pm}(a)$ and $A_{L\pm}(a)$ are functions of the black hole spin $a$, 
the specific forms of which are given in App.\ \ref{appendix:A}. 
Note that the zeroth order of the solutions $x_{m\pm}$ and $L_{m\pm}$ is exactly the results $r_c^{(2),(1)}$ and $\ell_b(r_c^{(2),(1)})$ for the Kerr case in 
\cite{Compere:2021bkk}\footnote{ In the Kerr case \cite{Compere:2021bkk}, their $r_c^{(1),(2)}$ is obtained by solving $E=1$, and therefore corresponds to the zeroth order of our results $x_{m\pm}$. Note that their upper index $(2)$ corresponds to $+$ sign in $x_{m\pm}$, and the upper index $(1)$ corresponds to $-$ in our notation. So does $\ell_b(r_c^{(2),(1)})$, which corresponds to the zeroth order of the MCO angular momentum $L_{m\pm}$. }. When $a=0$, and replace back $r$ instead of $x$, the solutions back to the Schwarzschild-like case\cite{Tu:2023xab} given by
\bea
r_{m\pm}&=&4-4A_\lambda+O(A_\lambda^2),\\
L_{m\pm}&=&\pm(4-\frac{9}{4}A_\lambda)+O(A_\lambda^2).
\eea

\begin{figure}[!ht]
     \centering
     \begin{subfigure}[b]{0.45\textwidth}
         \centering
         \includegraphics[width=\textwidth]{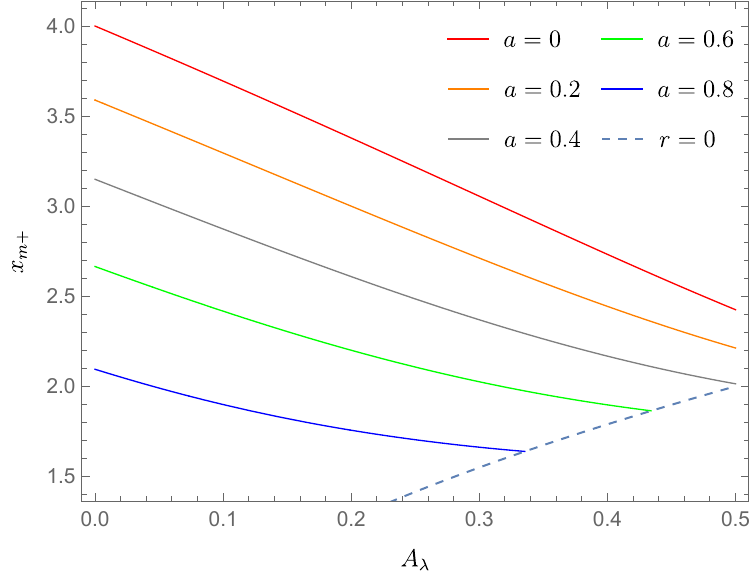}
         \caption{}
         \label{fig:mcoa}
     \end{subfigure}
     %\hfill
     \begin{subfigure}[b]{0.45\textwidth}
         \centering
         \includegraphics[width=\textwidth]{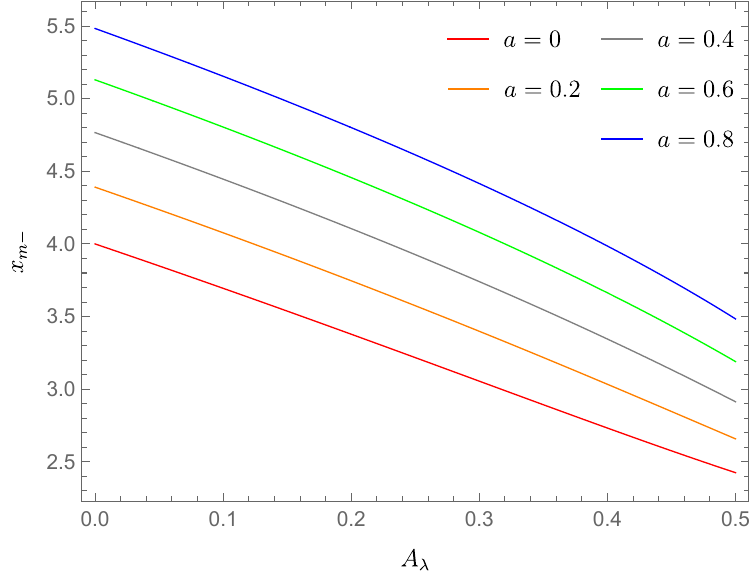}
         \caption{}
         \label{fig:mcoc}
     \end{subfigure}
     %\hfill
     \begin{subfigure}[b]{0.45\textwidth}
         \centering
         \includegraphics[width=\textwidth]{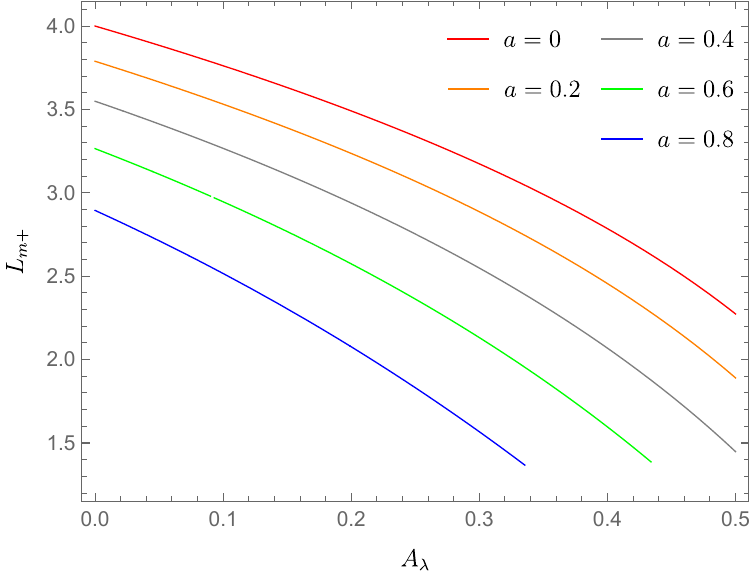}
         \caption{}
         \label{fig:mcob}
     \end{subfigure}
     %\hfill
     \begin{subfigure}[b]{0.45\textwidth}
         \centering
         \includegraphics[width=\textwidth]{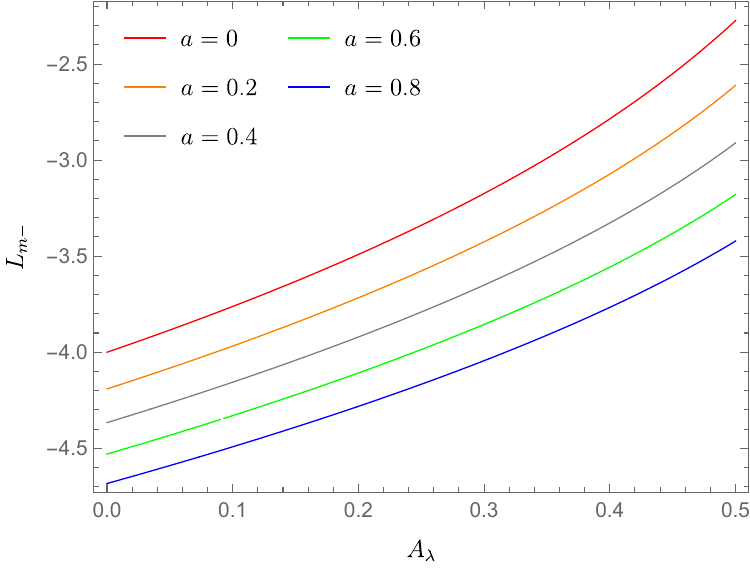}
         \caption{}
         \label{fig:mcod}
     \end{subfigure}
      \captionsetup{width=.9\textwidth}
       \caption{(a) and (c): MCO radius and angular momentum as functions of $ A_\lambda $ for prograde orbits. (b) and (d): MCO for retrograde orbits. The dashed line represents the position of the transition surface.}
        \label{fig:mco}
\end{figure}

In Fig.\ \ref{fig:mco}, we show the orbital radius and angular momentum of the MCO as functions of the quantum correction term $A_\lambda$. 

The dashed line in Fig.\ \ref{fig:mcoa} represents that the MCO is located at the position of the transition surface $x = \sqrt{8 A_\lambda}$. 
As $A_\lambda$ increases, both the MCO radius and the absolute value of angular momentum decrease, exhibiting a behavior consistent with that observed in the Schwarzschild-like LQG black hole. 
When the selected values of $(a, A_\lambda)$ enter region I as shown in Fig.\ \ref{fig:horizon}, the primary rotating object turns into a rotating wormhole. 
The curves denoting the position and angular momentum of prograde MCO intersect with the transition surface $r=0$, which implies that the prograde MCO could lie on the wormhole throat as the parameter $A_\lambda$ increases. 
However, the retrograde MCO would never reach the wormhole throat.

%%%%%%%%%%%%%%%%%%%%%%%%%%%%%%%%%%%%%%%%%%%%%%%%%%%%%%%%%%%%%%%%%%%%%%%%%%%%%%%%%%%%%%%%
\subsection{Properties of the Innermost Stable Circular Orbit}
%%%%%%%%%%%%%%%%%%%%%%%%%%%%%%%%%%%%%%%%%%%%%%%%%%%%%%%%%%%%%%%%%%%%%%%%%%%%%%%%%%%%%%%%

\begin{figure}[H]
     \centering
     \begin{subfigure}[b]{0.44\textwidth}
         \centering
         \includegraphics[width=\textwidth]{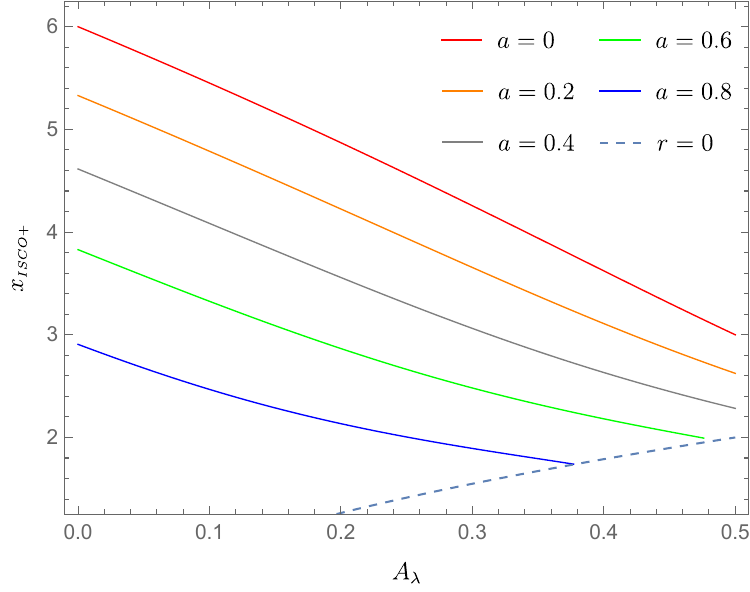}
         \caption{}
         \label{fig:iscoa}
     \end{subfigure}
     %\hfill
     \begin{subfigure}[b]{0.44\textwidth}
         \centering
         \includegraphics[width=\textwidth]{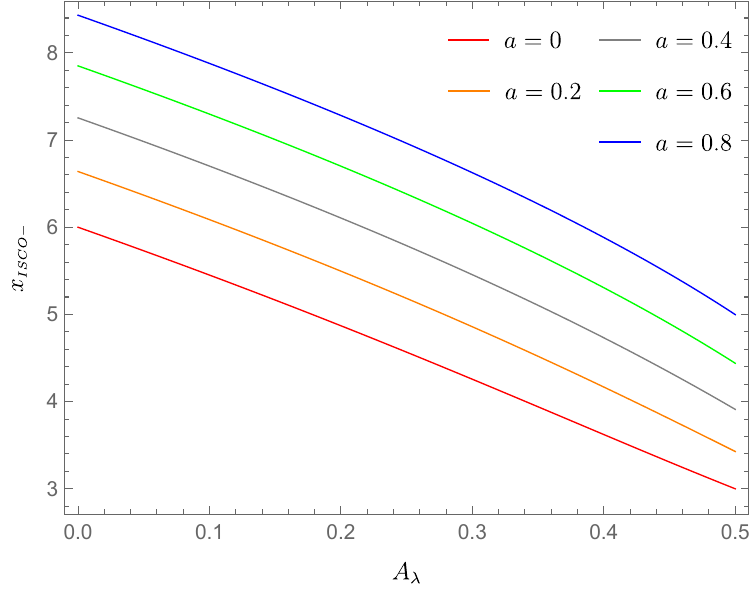}
         \caption{}
         \label{fig:iscob}
     \end{subfigure}
     %\hfill
     \begin{subfigure}[b]{0.44\textwidth}
         \centering
         \includegraphics[width=\textwidth]{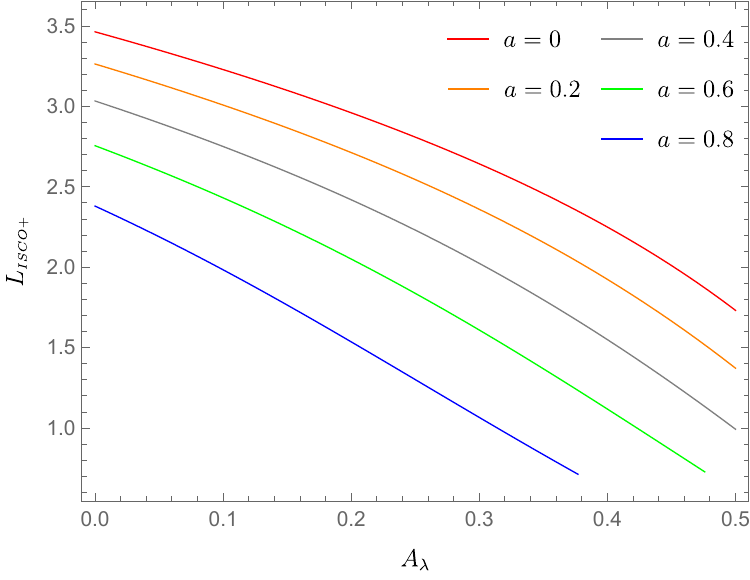}
         \caption{}
        \label{fig:iscoc}
     \end{subfigure}
     %\hfill
     \begin{subfigure}[b]{0.44\textwidth}
         \centering
         \includegraphics[width=\textwidth]{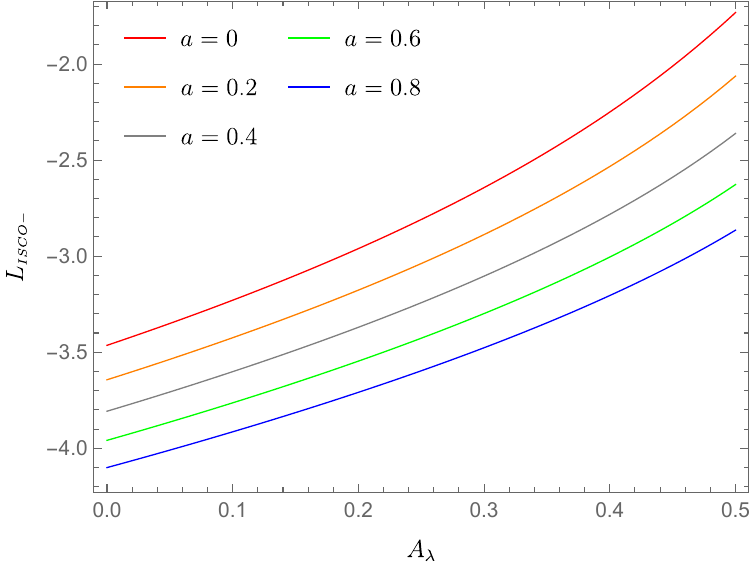}
         \caption{}
         \label{fig:iscod}
     \end{subfigure}
     %\hfill
     \begin{subfigure}[b]{0.44\textwidth}
         \centering
         \includegraphics[width=\textwidth]{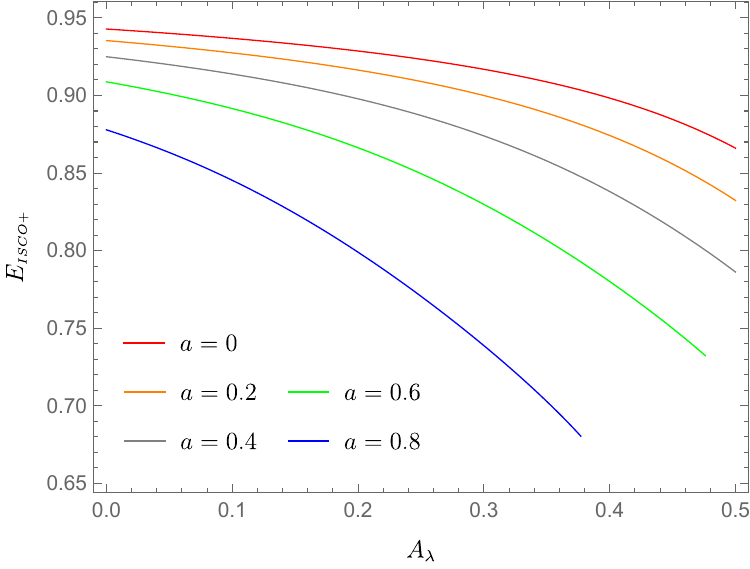}
         \caption{}
         \label{fig:iscoe}
     \end{subfigure}
     %\hfill
     \begin{subfigure}[b]{0.44\textwidth}
         \centering
         \includegraphics[width=\textwidth]{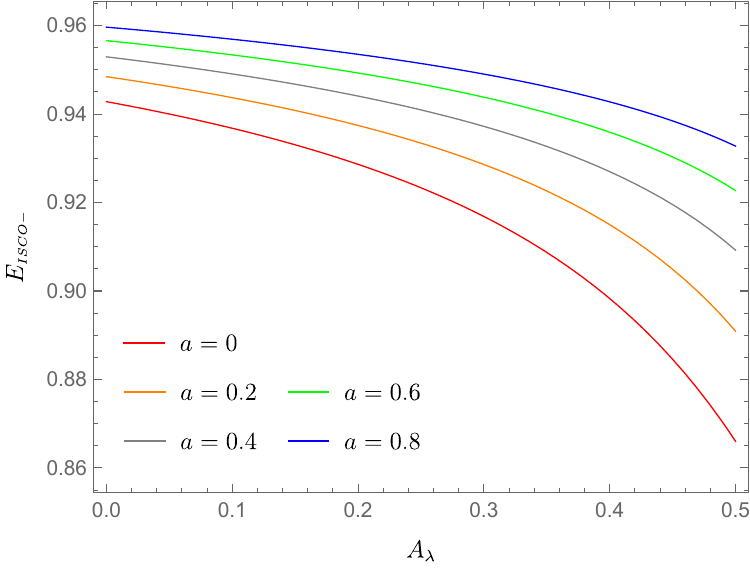}
         \caption{}
         \label{fig:iscof}
     \end{subfigure}
      \captionsetup{width=.9\textwidth}
       \caption{The position, angular momentum, and energy of prograde ISCO (left) and retrograde ISCO (right) as functions of $ A_\lambda $ with different values of $a$.}
        \label{fig:isco}
\end{figure}

The ISCO is a significant type of orbit, representing the circular orbital motion with a minimum value of angular momentum $L$ and energy $E$, 
and acting as the critical boundary between stable and unstable circular orbits. 
Combining the Eqs.\ \eqref{eq:Ecir} and \eqref{eq:Lcir} governing generic circular motion with the condition $R''(x) = 0$, the orbital radius $x_{ISCO\pm}$, angular momentum $L_{ISCO\pm}$, and energy $E_{ISCO\pm}$ of the prograde and retrograde ISCO can be numerically obtained, and are shown in Fig. \ref{fig:isco} as functions of $A_\lambda$ by fixing various values $a$, which denotes the spin of the primary rotating object.

As $A_\lambda$ increases, Fig. \ref{fig:isco} shows decreasing behaviors of the orbital radius, the absolute value of angular momentum, and the energy of both prograde and retrograde ISCO.

By comparing the prograde and retrograde ISCO motion, we observe that the absolute value of angular momentum and energy of prograde ISCO decreases much faster than that of retrograde ISCO as $A_\lambda$ increases. Since the prograde ISCO is much closer to the black hole where the gravitational field is much stronger, therefore, this different decreasing rate behavior suggests that the quantum corrections become more pronounced in the strong gravitational field. Note that as $a$ increases, these three ISCO quantities show similar behavior compared with the Kerr case.

However, an interesting phenomenon appears when the primary object becomes a wormhole as $A_\lambda$ increases. For larger values of $a$ where the parameters $ (a, A_\lambda) $ are located at the region I of Fig.\ \ref{fig:horizon}, the prograde ISCO radius curves $x_{ISCO+}$ intersect with the dashed line denoting the transition surface of the wormhole. This behavior is consistent with that of the MCO, suggesting that the prograde ISCO could lie on the transition surface located at the wormhole throat, while the retrograde ISCO could never touch. This phenomenon indicates that there will be no unstable prograde circular orbits in the rotating polymer wormhole universe with proper $ (a, A_\lambda) $. Such that once there exist unstable circular orbits, they must be in retrograde motion. 

\section{Analysis of Periodic Orbits and Rational Structures}
\label{sec4}

In this section, we conduct a detailed discussion on the performance of periodic orbits around a rotating polymer black hole. A periodic orbit, or a bound orbit, is a trajectory that can return to its initial state within a finite amount of time. It plays a crucial role in determining the motion trajectories of objects, which greatly facilitates the study and analysis of the evolution of EMRIs. Hence, it is valuable to investigate periodic orbital motion. To examine the effect of $A_\lambda$ on periodic orbits, we fix the black hole spin $a = 0.5$ for all subsequent calculations and plots in this section.

For a periodic orbit, the allowed ranges of energy and angular momentum are shown in Fig.\ \ref{fig:allowall}, where $x_p$ and $x_a$ represent the pericenter and apocenter of the root periodic orbit, respectively, satisfying 
\bea
R(x_p) = 0, \qquad R(x_a) = 0.
\eea

\begin{figure}[!ht]
     \centering
     \begin{subfigure}[b]{0.45\textwidth}
         \centering
         \includegraphics[width=\textwidth]{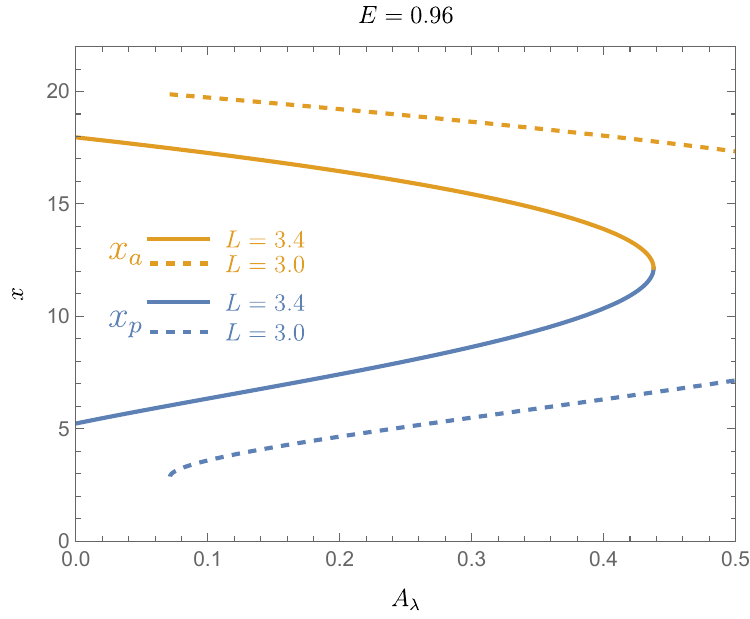}
         \caption{}
         %\label{fig:}
     \end{subfigure}
     %\hfill
     \begin{subfigure}[b]{0.45\textwidth}
         \centering
         \includegraphics[width=\textwidth]{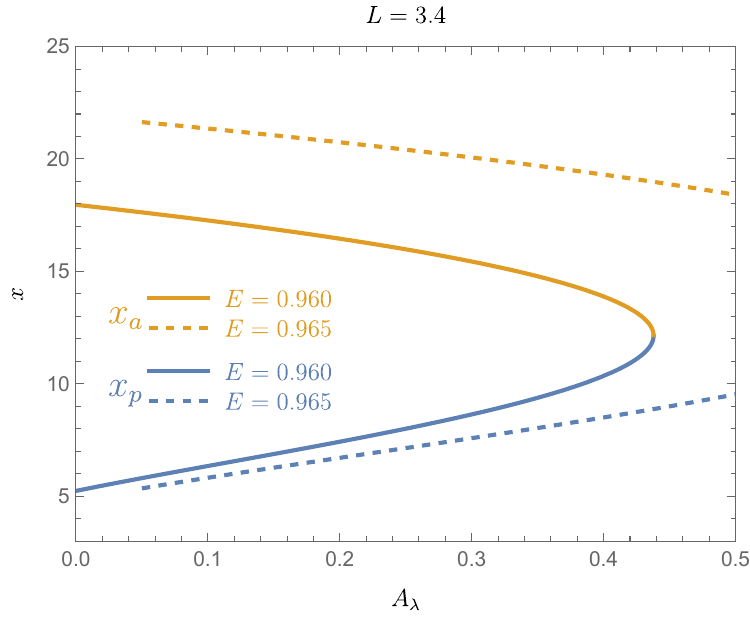}
         \caption{}
         %\label{fig:}
     \end{subfigure}
      \captionsetup{width=.9\textwidth}
       \caption{Turning points as a function of $ A_\lambda $, with fixed energy and angular momentum.}
       \label{fig:turn}
\end{figure}

In Fig.\ \ref{fig:turn}, we show the turning points of the bound orbits as functions of $A_\lambda$. 
We observe that as $A_\lambda$ increases, the apocenter of periodic orbits decreases, while the pericenter increases. 
When the apocenter and pericenter coincide, they correspond to stable circular orbits. 
As the orbital energy increases or the absolute value of the angular momentum decreases, the apocenter increases while the pericenter decreases for a given $A_\lambda$. In addition, Fig.~6(a) (fixed energy) shows that for larger $A_\lambda$, a smaller increase in angular momentum is required for an elliptical orbit to transition into a circular one through gravitational radiation.  
A similar trend appears in Fig.~6(b) (fixed angular momentum), where larger $A_\lambda$ require a smaller loss of energy for the transition.

For example, in Fig.~6(b), suppose that the solid lines representing the apocenter and pericenter for an orbital energy $E=0.960$ coincide at $A_{\lambda1}$.  
In a universe with $A_\lambda = A_{\lambda1}$, a periodic orbit with energy $E = 0.965$, denoted by the dashed line, must radiate about $0.005$ of energy to become circular.  
By contrast, for periodic motion with $A_{\lambda2} < A_{\lambda1}$, the apocenter and pericenter merge at $A_{\lambda2}$, which lies just inside the solid lines and therefore corresponds to an energy slightly below $E = 0.960$.  
In this case, the periodic orbit must radiate more energy than in the $A_{\lambda1}$ scenario.  
From this comparison, we conclude that larger $A_\lambda$ values lead to less energy loss when a periodic orbit evolves toward a quasi-circular configuration. 
This behavior is directly relevant to EMRIs, where both $E$ and $L$ evolve due to gravitational radiation.  
The reduced energy loss at larger $A_\lambda$ could therefore influence the inspiral dynamics and gravitational wave signatures, providing a potential observational test of this quantum-gravity scenario.

Additionally, for fixed orbital energy and angular momentum, increasing $A_\lambda$ can modify the root structure of the radial equation, thereby transforming a bound orbit into a circular one. However, within certain ranges of $(E, L)$, the apocenter and pericenter cease to coincide within the valid range of $A_\lambda$, and the circular motion does not exist in these cases.

Note that there are no pericenter and apocenter points for small values of $A_\lambda$, which is also shown in Fig.\ \ref{fig:RadialA}, indicating the absence of periodic orbits due to the single-root structure of radial potential with selected parameters. 
However, as $A_\lambda$ increases, the radial potential evolves into a three-root structure, and periodic orbits start to appear. 
In this case, the second root of the radial potential corresponds to the pericenter of the periodic orbit, while the third root corresponds to the apocenter.

\begin{figure}[!ht]
     \centering
     \begin{subfigure}[b]{0.45\textwidth}
         \centering
         \includegraphics[width=\textwidth]{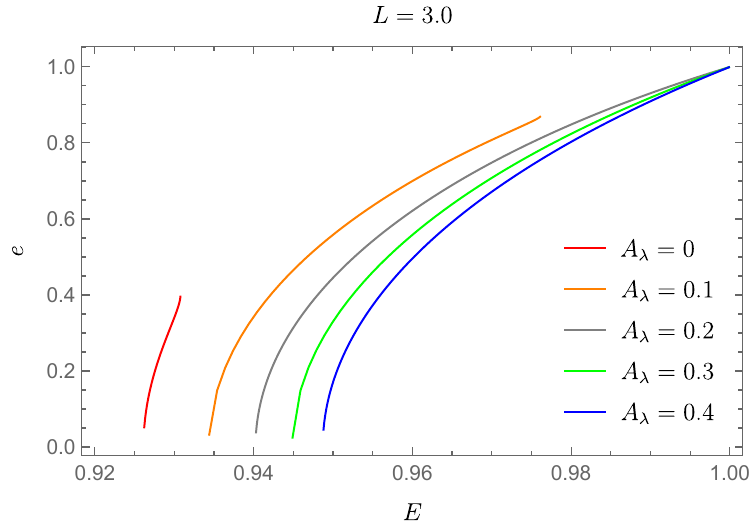}
         \caption{}
         \label{fig:evAa}
     \end{subfigure}
     %\hfill
     \begin{subfigure}[b]{0.45\textwidth}
         \centering
         \includegraphics[width=\textwidth]{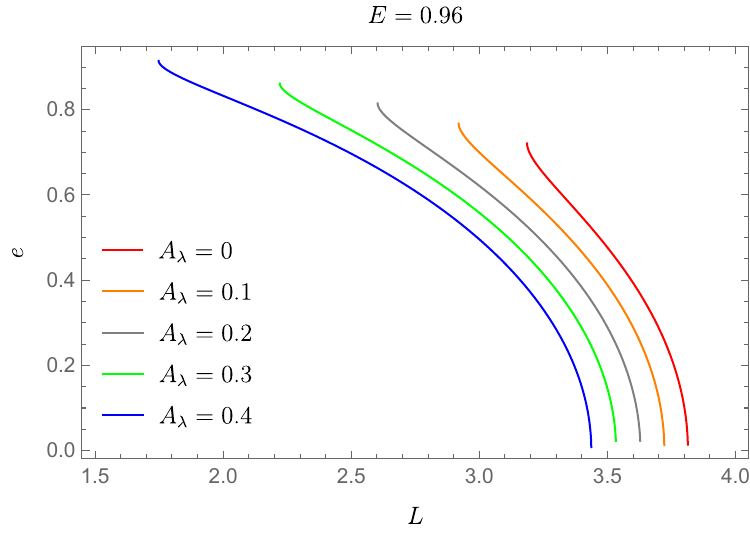}
         \caption{}
         \label{fig:evAb}
     \end{subfigure}
     %\hfill
     \begin{subfigure}[b]{0.45\textwidth}
         \centering
         \includegraphics[width=\textwidth]{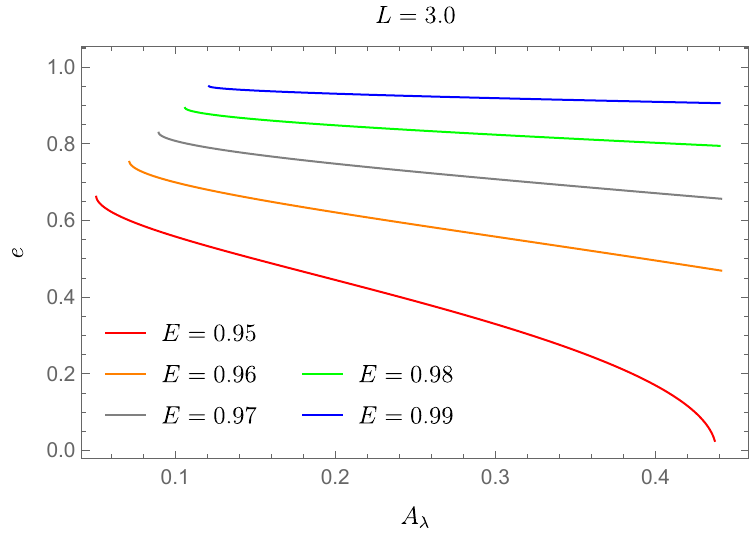}
         \caption{}
         \label{fig:evAc}
     \end{subfigure}
     %\hfill
     \begin{subfigure}[b]{0.45\textwidth}
         \centering
         \includegraphics[width=\textwidth]{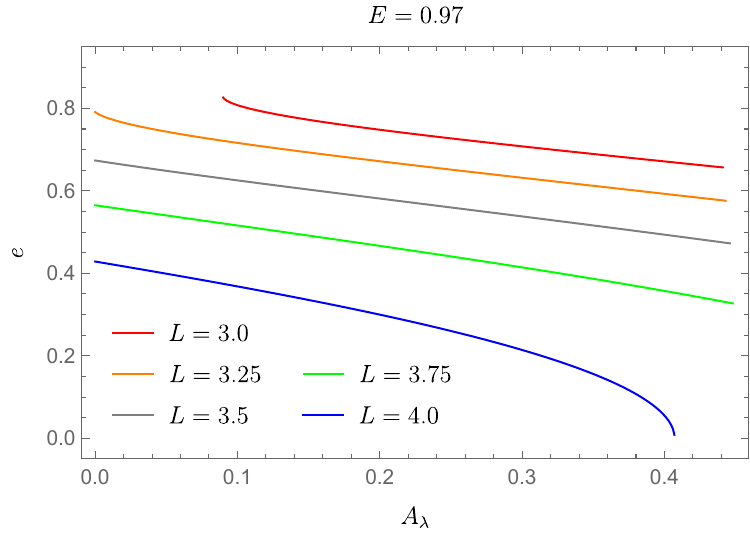}
         \caption{}
         \label{fig:evAd}
     \end{subfigure}
      \captionsetup{width=.9\textwidth}
       \caption{The variation of eccentricity due to energy $E$, angular momentum $L$, and the parameter $A_\lambda$.}
        \label{fig:evA}
\end{figure}

The eccentricity of the bound motion is defined by the apocenter and pericenter as
\bea
e = \frac{x_a - x_p}{x_a + x_p}.
\eea
In Fig.\ \ref{fig:evA}, we show the eccentricity in terms of $L$, $E$, and $A_\lambda$. 

The eccentricity increases as energy increases and decreases as the angular momentum increases. 
For the red and orange curves with lower parameters $A_\lambda=(0, 0.1)$ in Fig.\  \ref{fig:evAa}, the eccentricity can not take the whole range. 
However, the gray, green, and blue curves with higher parameters $A_\lambda=(0.2, 0.3, 0.4)$ could take the whole range, and coincide at $ e = 1 $ as the energy approaches $ E = 1 $, which implies that the apocenter approaches infinity at this marginal point with $E=1$. 
Similarly, in Fig.\ \ref{fig:evAb}, as $A_\lambda$ increases, the range of angular momentum for periodic orbits narrows. 
That is because, for small $A_\lambda$ cases, the bound motion is limited to a small energy range in the $(E, L)$ space as shown in Fig.\ \ref{fig:allowall}. 
In these cases, the eccentricity varies from $e=0$ to $e=1$, which corresponds to the starting separation of the double roots denoting the stable and unstable circular motion, respectively. 
As $A_\lambda$ increases, the allowed region in the  $ (E, L) $ space for bound motion changes. 
Such that the energy and angular momentum could take values in a wider range, and therefore affect the range of eccentricity for the bound motion.

In Figs.\ \ref{fig:evAc} and \ref{fig:evAd}, we further investigate the effect of $A_\lambda$ on eccentricity. The left limit points of all the curves in Figs.\ \ref{fig:evAc} and \ref{fig:evAd} correspond to the homoclinic orbits, with which the pericenter coincides with the first single root of the radial potential, and as a result form the structure of unstable circular motion. These two sub-figures show that eccentricity decreases as $A_\lambda$ increases. Besides, the effect of $A_\lambda$ on eccentricity is weaker for the bound motion with higher energy and is stronger for the motion with higher angular momentum.

In Ref.\ \cite{Levin:2008mq}, a relationship between rational numbers and periodic orbits was proposed. Specifically, a triplet of integers $ (z, w, v) $ was introduced to characterize periodic orbits around black holes, denoting the zoom, whirl, and vertex behaviors, respectively. This triplet is related to rational numbers given by
\bea
q_\pm &\equiv& \frac{w_\varphi}{w_r} \mp 1 = \frac{\Delta\varphi_{r\pm}}{2 \pi} \mp1 = \pm (w +\frac{v}{z} ),
\eea
where $w_r$ and $w_\varphi$ denote the radial and angular frequencies of the periodic orbit, respectively. $\Delta \varphi_{r\pm}$ represents the equatorial angle over one radial period, and the $ \pm $ signs correspond to prograde and retrograde periodic orbits. We supplement the relationship between rational numbers and retrograde periodic orbits, with additional details provided in App.\ \ref{appendix:B}. The quantity $q$ can be calculated by
\bea
q_\pm &=& \frac{1}{\pi}  \int_{x_p}^{x_a} \frac{\dot{\phi}}{\dot{x}}  \d x  \mp 1
\nonumber\\
  &=& \frac{1}{\pi}  \int_{x_p}^{x_a} \frac{L ( x - 2 )x + 2 a E ( x - 3 A_\lambda )}{(a^2 + (x - 2)x)\sqrt{R(x)}} \frac{x}{\sqrt{x^2 - 8 A_\lambda}} dx  \mp 1.
\eea

\begin{figure}[!ht]
     \centering
     \begin{subfigure}[b]{0.45\textwidth}
         \centering
         \includegraphics[width=\textwidth]{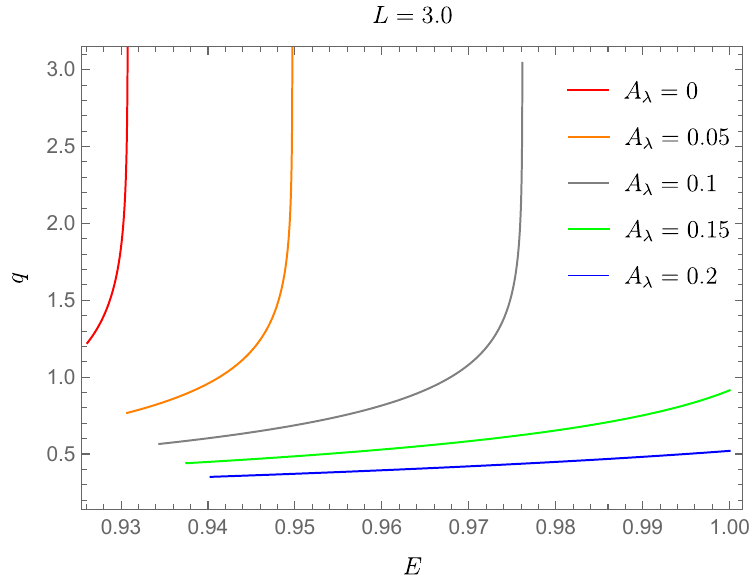}
         \caption{}
         \label{fig:qvELa}
     \end{subfigure}
     %\hfill
     \begin{subfigure}[b]{0.45\textwidth}
         \centering
         \includegraphics[width=\textwidth]{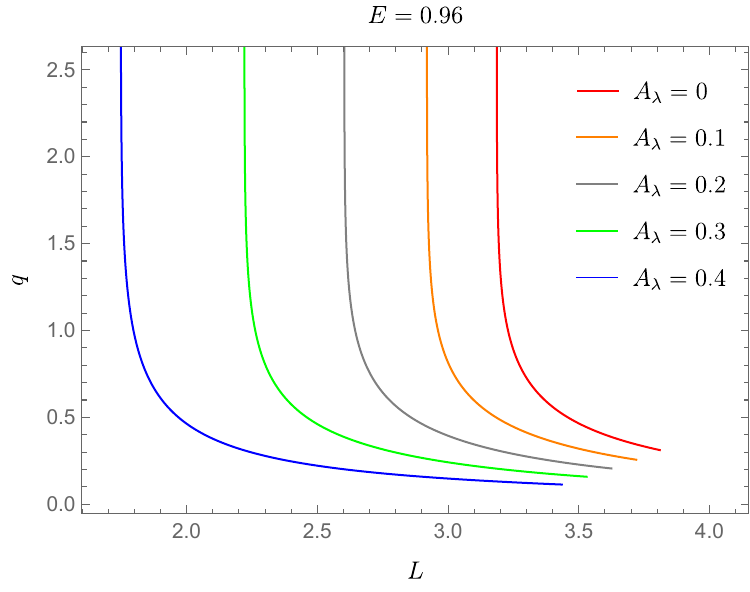}
         \caption{}
         \label{fig:qvELb}
     \end{subfigure}
     %\hfill
     \begin{subfigure}[b]{0.45\textwidth}
         \centering
         \includegraphics[width=\textwidth]{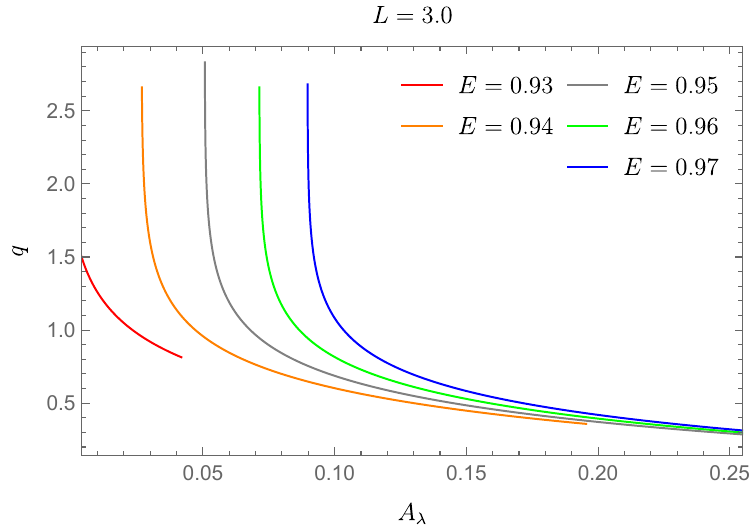}
         \caption{}
         \label{fig:fix-L-and-Ea}
     \end{subfigure}
     %\hfill
     \begin{subfigure}[b]{0.45\textwidth}
         \centering
         \includegraphics[width=\textwidth]{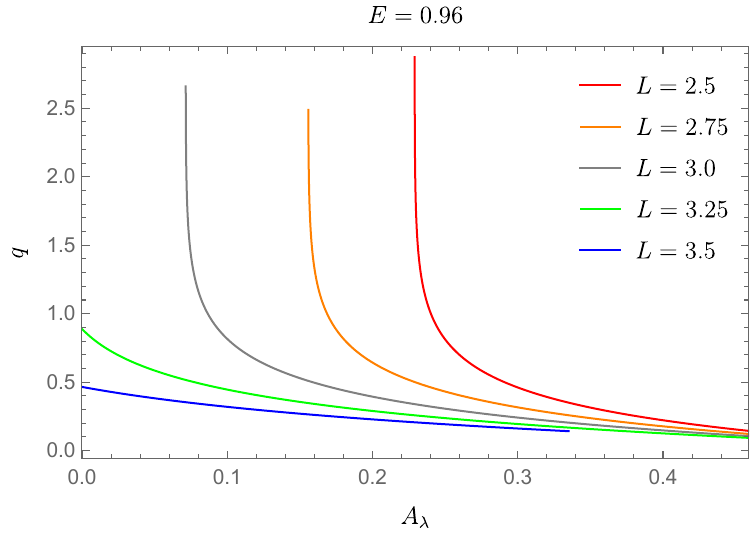}
         \caption{}
         \label{fig:fix-L-and-Eb}
     \end{subfigure}          
      \captionsetup{width=.9\textwidth}
       \caption{The rational number $q$ as a function of energy $E$, angular momentum $L$, and the parameter $A_\lambda$.}
        \label{fig:qvEL}
\end{figure}

In Fig.\ \ref{fig:qvEL}, we show the rational number $q$ as a function of energy $E$, angular momentum $L$, and the parameter $A_\lambda$. 
For the green and blue curves in Fig.\  \ref{fig:qvELa}, terminate at $ E = 1 $, where $ q $ remains finite. 
While for the red, orange, and gray curves, $ q $ increases slowly at first, and then exhibits rapid, explosive growth as $ E $ approaches its maximum value. 
The divergence point of $q$ corresponds to the case that the pericenter coincides with the first single root and, in turn with the unstable circular orbit. In contrast, the minimum value of $q$ represents the rational number associated with a stable circular orbit, as discussed in Ref.\ \cite{Levin:2008mq}. 
As $ A_\lambda $ increases, the divergence point of $ q $ shifts to the right but remains truncated at $ E = 1 $. Beyond $ E = 1 $, the particle escapes to infinity, rendering periodic orbits non-existent. Furthermore, based on the allowed region of the bound motion shown in Fig.\ \ref{fig:allowall}, we observe that $ q $ exhibits a divergence point only when the orbital angular momentum satisfies $L < L_m$.

Next, we fix the energy and plot $q$ as a function of angular momentum $L$ for prograde orbits with different values of $A_\lambda$ in Fig.\ \ref{fig:qvELb}. 
In this case, $q$ diverges when $ L $ reaches its minimum value, and it decreases rapidly as $L$ increases near this point. 
Additionally, the divergence point shifts to the left as $ A_\lambda $ increases, with the magnitude of the shift becoming progressively larger. 
Similar to Fig.\ \ref{fig:qvELa},  the divergence point and the minimum point of the curves in Fig.\ \ref{fig:qvELb} correspond to the unstable circular motion and stable circular motion, respectively. 
However, unlike Fig.\ \ref{fig:qvELa}, all curves in Fig.\ \ref{fig:qvELb} exhibit a divergence point. 
That is because, as shown in Fig.\ \ref{fig:allowall}, both stable and unstable circular orbits are always present within the allowed range of $L$ for this fixed energy.

We further investigate the effect of the parameter $A_\lambda$ on the rational number $q$. 
We observe that $q$ decreases gradually as $A_\lambda$ increases, but exhibits a sharp drop near the divergence point. 
This behavior is analogous to the variation of $q$ with angular momentum, as shown in Fig.\ \ref{fig:qvELb}. 
In Fig.\  \ref{fig:fix-L-and-Ea}, we see that the divergence point shifts to the left as energy decreases, while in Fig.\ \ref{fig:fix-L-and-Eb}, the divergence point shifts leftward as angular momentum increases. 
The curves terminating at $A_\lambda = 0$, indicate that for smaller energy values and larger angular momenta, $q$ does not diverge within the range of $ A_\lambda$.

For periodic orbits characterized by integer triplets $ (z, w, v) $, we numerically computed the angular momentum $L$ corresponding to a specific rational number $q$ for prograde orbits, varying the parameter $A_\lambda$ while fixing $ E = 0.96 $. 
Similarly, we calculated the energy $E$ for a fixed angular momentum $ L = 3.05 $. The results are summarized in Tabs. \ref{table:energy} and \ref{table:angular}. 
\begin{table}[h!t]
\centering
\begin{tabular}{c c c c c c c c c}
\hline
\hline
$A_\lambda$ & $E(1,1,0)$ & $E(1,2,0)$ & $E(2,1,1)$ & $E(2,2,1)$ & $E(3,1,2)$ & $E(3,2,2)$ & $E(4,1,3)$ & $E(4,2,3)$ \\
\hline
0 & 0.929596 & 0.937801 & 0.936415 & 0.93811 & 0.937122 & 0.938145 & 0.93736 & 0.938158 \\
\hline
0.02 & 0.937226 & 0.945209 & 0.943924 & 0.945474 & 0.94459 & 0.945502 & 0.94481 & 0.945512 \\
\hline
0.04 & 0.945817 & 0.953663 & 0.952451 & 0.953898 & 0.953087 & 0.953922 & 0.953295 & 0.95393 \\
\hline
0.06 & 0.955458 & 0.96324 & 0.962078 & 0.955458 & 0.962693 & 0.963475 & 0.962892 & 0.963482 \\
\hline
0.08 & 0.966248 & 0.974031 & 0.972901 & 0.974231 & 0.973504 & 0.97425 & 0.973697 & 0.974256 \\
\hline
0.1 & 0.978302 & 0.986149 & 0.985035 & 0.98634 & 0.985634 & 0.986358 & 0.985824 & 0.986363 \\
\hline
\hline
\end{tabular}
\captionsetup{width=.9\textwidth}
\caption{The energy $E$ for the periodic orbits with different $(z,w,v)$ and different parameter $A_{\lambda}$, where we fix the angular momentum parameter $L = 3.05$. }
\label{table:energy}
\end{table}
\begin{table}[h!t]
\centering
\begin{tabular}{c c c c c c c c c}
\hline
\hline
$A_\lambda$ & $L(1,1,0)$ & $L(1,2,0)$ & $L(2,1,1)$ & $L(2,2,1)$ & $L(3,1,2)$ & $L(3,2,2)$ & $L(4,1,3)$ & $L(4,2,3)$ \\
\hline
0 & 3.22941 & 3.18767 & 3.19410 & 3.18645 & 3.19072 & 3.18633 & 3.18961 & 3.18629 \\
\hline
0.02 & 3.17910 & 3.13751 & 3.14393 & 3.13630 & 3.14055 & 3.13618 & 3.13945 & 3.13614 \\
\hline
0.04 & 3.12725 & 3.08582 & 3.09221 & 3.08460 & 3.08885 & 3.08448 & 3.08775 & 3.08444 \\
\hline
0.06 & 3.07377 & 3.03249 & 3.03887 & 3.03128 & 3.03552 & 3.03116 & 3.03442 & 3.03112 \\
\hline
0.08 & 3.01857 & 2.97744 & 2.98381 & 2.97623 & 2.98046 & 2.97611 & 2.97937 & 2.97607 \\
\hline
0.1 & 2.96154 & 2.92055 & 2.92691 & 2.91934 & 2.92357 & 2.91922 & 2.92248 & 2.91918 \\
\hline
\hline
\end{tabular}
\captionsetup{width=.9\textwidth}
\caption{The angular momentum parameter $L$ for the periodic orbits with different $(z,w,v)$ and different parameter $A_{\lambda}$, where we fix the energy $E = 0.96$.}
\label{table:angular}
\end{table}
Our findings show that the energy of the test particle in periodic orbits increases with $A_\lambda$, while the angular momentum decreases as $A_\lambda$ increases. 
Compared to the classical Kerr black hole, periodic orbits around the rotating LQG polymer black hole exhibit lower energy and higher angular momentum for the same integer triplets $(z, w, v)$.

\begin{figure}[H]
     \centering
         \includegraphics[width=\textwidth]{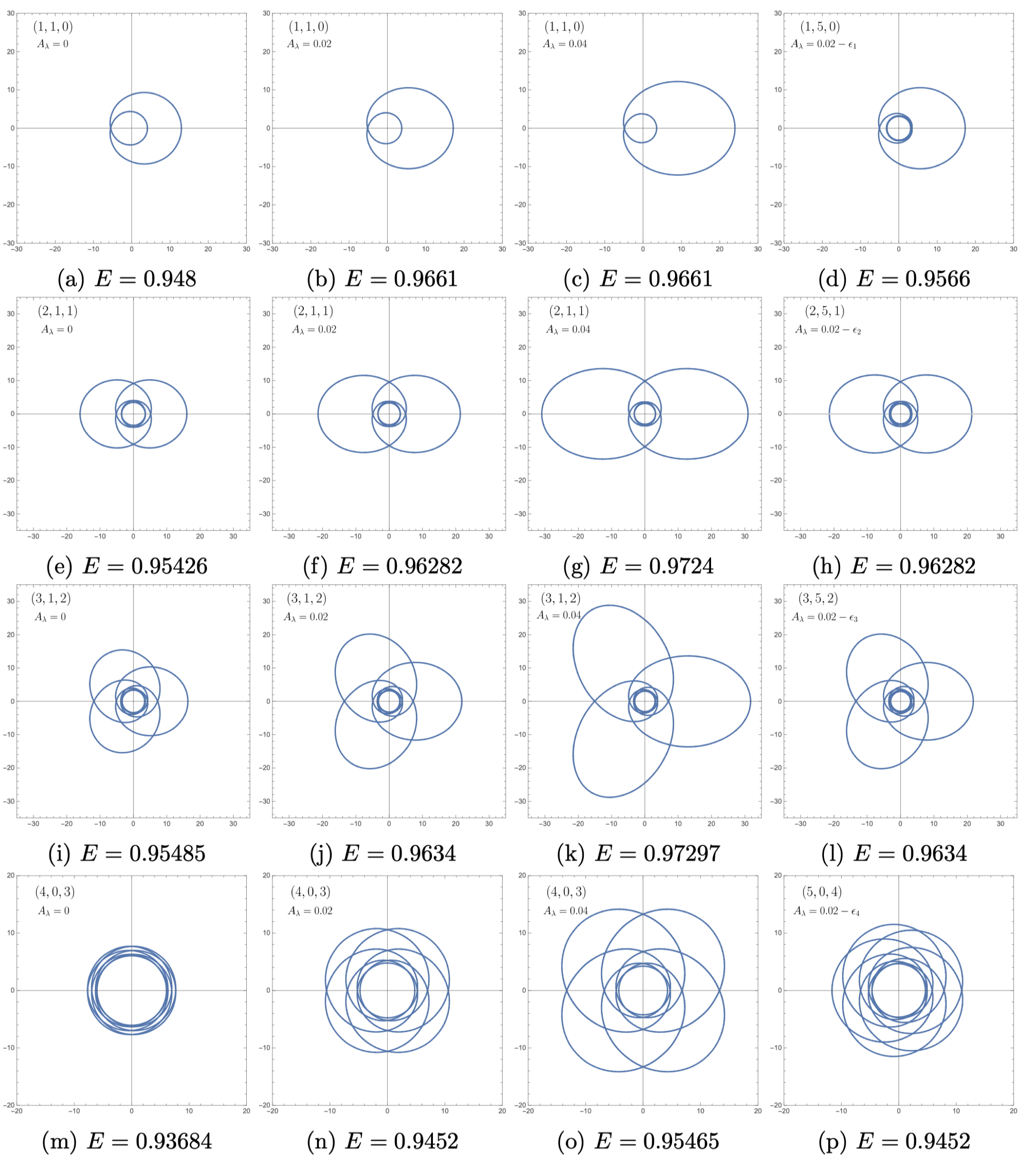}
 %\hfill   %%%%%%%%%%%%%%%%%%%%%%%%%%%%%%%%%%%%          
      \captionsetup{width=.9\textwidth}
       \caption{Periodic orbits with different $(z, w, v)$ around the rotating LQG black hole with $L = 3.16$, where $\epsilon_1 = 0.0177$, $\epsilon_2 = 0.003$, $\epsilon_3 = 0.0017$, $\epsilon_4 = 0.0075$.}
        \label{fig:periodicL}
\end{figure}

We plot the periodic orbits for different indices $ (z, w, v) $ with a fixed angular momentum $ L = 3.16 $ in Fig.\ \ref{fig:periodicL}. 
Comparing the first three columns, we observe that as $ A_\lambda $ increases, the energy also increases. 
This leads to an increase in the apocenter, a decrease in the pericenter, and an overall expansion of the orbit. 
As shown in Fig.\ \ref{fig:evA}, the eccentricity of the periodic orbit increases with energy. 
Although the eccentricity decreases with $ A_\lambda $, this effect is less pronounced than the influence of energy, resulting in an overall increase in the eccentricity of the periodic orbit.

\begin{figure}[H]
     \centering
         \includegraphics[width=\textwidth]{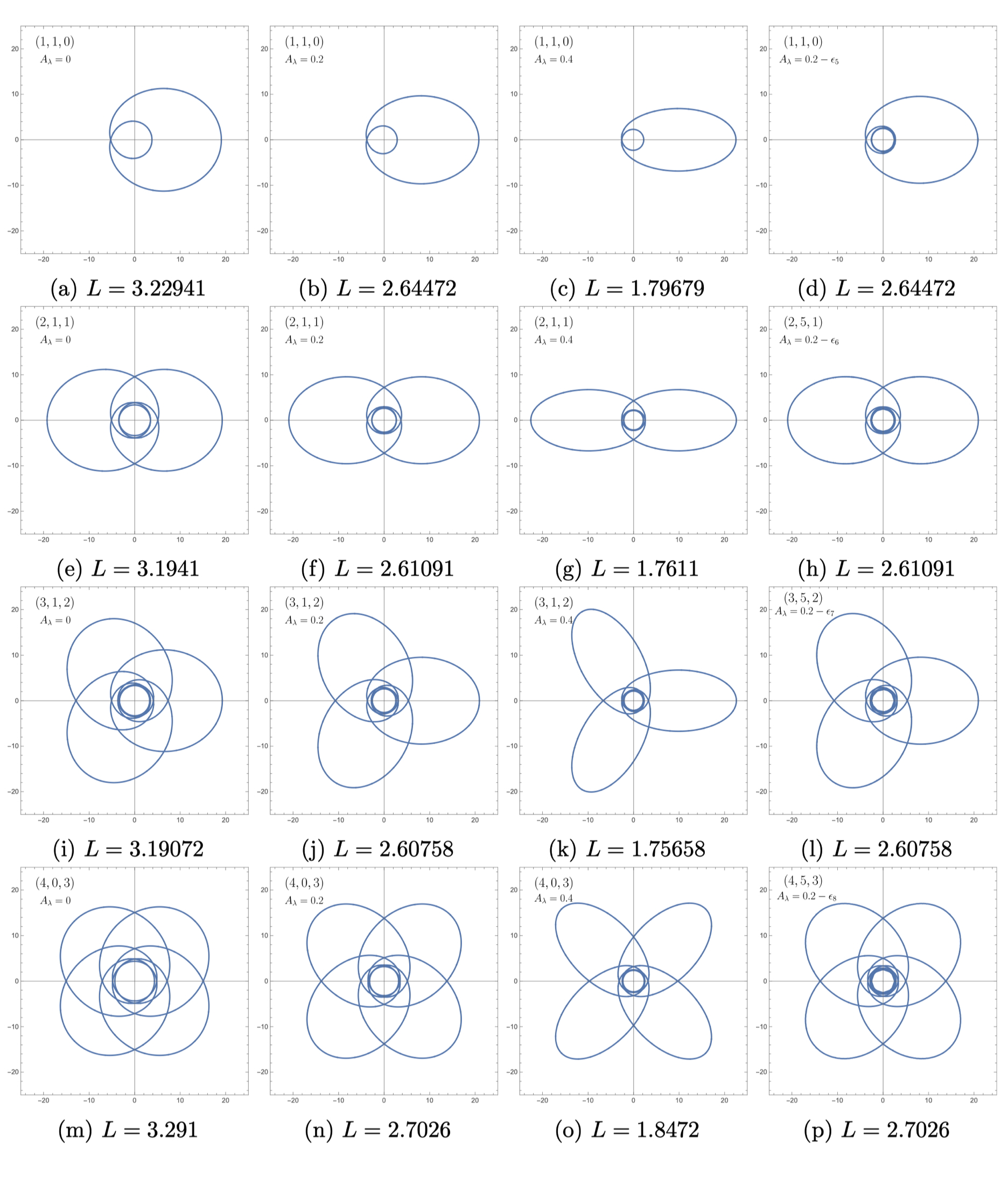}
 %\hfill   %%%%%%%%%%%%%%%%%%%%%%%%%%%%%%%%%%%%          
      \captionsetup{width=.9\textwidth}
       \caption{Periodic orbits with different $(z, w, v)$ around the rotating LQG black hole with $E = 0.96$, where $\epsilon_5 = 0.0122$, $\epsilon_6 = 0.0023$, $\epsilon_7 = 0.0013$, $\epsilon_8 = 0.02961$.}
        \label{fig:periodicE}
\end{figure}

Next, we fix the energy at $ E = 0.96 $ and plot the periodic orbit images for different indices $ (z, w, v) $ in Fig.\ \ref{fig:periodicE}. 
Comparing the first three columns, which correspond to the same indices $ (z, w, v) $, we find that increasing $ A_\lambda $ reduces the angular momentum, which in turn increases the apocenter and decreases the pericenter. 
Unlike Fig.\ \ref{fig:periodicL}, the periodic orbit gradually becomes more elongated without showing a significant expansion. 
Based on Fig.\ \ref{fig:evA}, the eccentricity increases as angular momentum decreases, and it also increases with $ A_\lambda $, both of which contribute to an overall increase in the eccentricity of the periodic orbit.

Finally, by comparing the second and fourth columns in Fig.\ \ref{fig:periodicL} with those in Fig.\ \ref{fig:periodicE}, where energy and angular momentum are fixed, we observe that when $ A_\lambda $ is decreased to very small values (as shown for small $ \epsilon $ in the figures), the rational number $ q $ increases rapidly. This is reflected in the fact that near the pericenter, the orbit undergoes many more revolutions. Notably, in the last plot of Fig.\ \ref{fig:periodicL}, for this specific $ (E, L) $ parameter set, the rational number does not diverge as $A_\lambda$ decreases. Instead, it reaches its maximum value when $ A_\lambda = 0 $, corresponding to the curve in Fig.\ \ref{fig:qvEL} that does not exhibit any divergence points. However, its rational number can reach 0.8, so we present the case for indices $ (5, 0, 4) $.

%%%%%%%%%%%%%%%%%%%%%%%%%%%%%%%%%%%%%%%%%%%%%%%%%%%%%
\section{Conclusion and Outlook}
\label{sec5}
%%%%%%%%%%%%%%%%%%%%%%%%%%%%%%%%%%%%%%%%%%%%%%%%%%%%%

In this paper, we have explored the geodesic motion of massive test particles in the polymer Kerr-like spacetime within the framework of loop quantum gravity, focusing on quantum-corrected effects represented by the parameter $A_\lambda$. 
An essential aspect of our findings involves the constraint on the orbital angular momentum, specifically discussed at the beginning of Sec.\ \ref{sec3}. 
We identified that as $A_\lambda$ increases, the range of permissible angular momentum for ingoing motion shrinks. 
This constraint affects the dynamics of inspiraling systems, requiring particles to undergo more orbits before reaching the plunging stage. 
In particular, quantum corrections reduce the angular momentum range for which plunging orbits are allowed, leading to more extensive orbital evolution before the particles can fall into the black hole.

Our analysis reveals that as $A_\lambda$ increases, both the radii and angular momenta of the ISCO and MCO decrease, indicating that quantum effects become more significant near the black hole where the gravitational field is strongest. 
This also results in a decrease in the ISCO energy for both prograde and retrograde orbits. 
Moreover, when the spacetime parameters $(a,A_\lambda)$ fall within specific ranges corresponding to a wormhole-like spacetime, we observe that the prograde ISCO and MCO could lie on the transition surface at the wormhole throat. 

Additionally, we investigated the behavior of periodic orbits and the associated rational number $q$. 
As $A_\lambda$ increases, the pericenter and apocenter of these orbits shift, with the eccentricity of the bound motion decreasing. 
Our analysis further suggests that larger $A_\lambda$ reduces energy loss and promotes circular motion.  
Such behavior could influence the inspiral dynamics and gravitational-wave signatures of EMRIs, offering a potential observational test of this quantum-gravity scenario.
The analysis of rational numbers also showed that quantum corrections cause a divergence in $q$, marking the transition between stable and unstable circular orbits, particularly for retrograde motion. 
This behavior has important implications for understanding the orbital dynamics in strong gravitational fields, such as those around black holes.

Our study provides valuable insights into the influence of quantum gravity on the motion of particles in black hole spacetimes, offering predictions that could be tested in future astrophysical observations, particularly those related to EMRIs. 
The work also paves the way for further research into the effects of quantum gravity in these systems, including the analysis of null geodesics, gravitational wave astronomy, and a more refined treatment of the angular momentum constraints.

\vspace{10pt}
{\noindent \bf Acknowledgments.}
%|--------------------------------------------------------------------|

We would like to appreciate Dr. Hyat Huang for helpful discussion. Y.\ L.\ is financially supported by Natural Science Foundation of Shandong Province under Grants No.\ ZR2023QA133 and Yantai University under Grants No.\ WL22B218. C.\ L.\ is supported by the National Natural Science Foundation of China under Grant No.\ 12175108, and Yantai University under Grant No.\ WL22B224.  

%|--------------------------------------------------------------------|

%%%%%%%%%%%%%%%%%%%%%%%%%%%%%%%%%%%%%%%%%%%%%%%%%%%%%%
\appendix
%%%%%%%%%%%%%%%%%%%%%%%%%%%%%%%%%%%%%%%%%%%%%%%%%%%%%%

%%%%%%%%%%%%%%%%%%%%%%%%%%%%%%%%%%%%%%%%%%%%%%%%%%%%%%
\section{Coefficients in the Perturbative Expansion of MCO Parameters}
\label{appendix:A}
%%%%%%%%%%%%%%%%%%%%%%%%%%%%%%%%%%%%%%%%%%%%%%%%%%%%%%

Here we provide the explicit expressions for $A_{x\pm}(a)$ and $A_{L\pm}(a)$ in Eqs.\ \eqref{eq:3.14} and \eqref{eq:3.15}. 
The coefficients $A_{x\pm}(a)$ and $A_{L\pm}(a)$ are given by
\bea
&&A_{x+}(a)=\frac{A1}{A2},\quad A_{x-}(a)=\frac{A3}{A4},\\
&&A_1=(-512 + 1344 a - 1232 a^2 + 456 a^3 - 57 a^4 + a^5)\sqrt{1-a}\nonumber\\
&&\qquad-512 + 1600 a - 1840 a^2 + 936 a^3 - 195 a^4 + 11 a^5,\\
&&A_2=(-512 + 1280 a - 1120 a^2 + 400 a^3 - 50 a^4 + a^5)\sqrt{1-a}\nonumber\\
&&\qquad+2 (-256 + 768 a - 848 a^2 + 416 a^3 - 85 a^4 + 5 a^5),\\
&&A_3=(512 + 1344 a + 1232 a^2 + 456 a^3 + 57 a^4 + a^5)\sqrt{1+a}\nonumber\\
&&\qquad+512 + 1600 a + 1840 a^2 + 936 a^3 + 195 a^4 + 11 a^5,\\
&&A_4=(512 + 1280 a + 1120 a^2 + 400 a^3 + 50 a^4 + a^5)\sqrt{1+a}\nonumber\\
&&\qquad+2 (256 + 768 a + 848 a^2 + 416 a^3 + 85 a^4 + 5 a^5),\\
&&A_{L+}=(2 \sqrt{1-a}-a  )   (\sqrt{-  ((a-1)   (a^2-4   (\sqrt{1-a}+2  ) a+8   (\sqrt{1-a}+1  )  )  )}   (-a^2 (4  A_{x+}\nonumber\\
&&\qquad+3) +a   (2   (5 \sqrt{1-a}+8  )  A_{x+}+6 \sqrt{1-a}+15  )-6   (2   (\sqrt{1-a}+1  )  A_{x+}+\sqrt{1-a}\nonumber\\
&&\qquad+2  )  ) +2 (a-1) a   (a-2   (\sqrt{1-a}+1  )  )  A_{x+}+6 (a-1) a   (a-2   (\sqrt{1-a}+1  )  )  ))/(1\nonumber\\
&&\qquad-a)   (a   (a+(2 (1-a)   (-2 a+4 \sqrt{1-a}+2  )   (  (\sqrt{1-a}+5  ) a^2-2   (5 \sqrt{1-a}+7  ) a\nonumber\\
&&\qquad-4 \sqrt{1-a}-6  )+4   (\sqrt{1-a}+1  )  )^2 + 8   (\sqrt{1-a}+1  )  )  A_{x+},\\
&&A_{L-}=\frac{3 (a^3 + 12 (1 + \sqrt{1+a}) + 3 a^2 (5 + 2 \sqrt{1+a}) + 
   a (26 + 20 \sqrt{1+a})) A_\lambda}{\sqrt{1+a} (a + 2 \sqrt{1+a}) (2 + a + 2 \sqrt{1+a})^2}.
\eea

%%%%%%%%%%%%%%%%%%%%%%%%%%%%%%%%%%%%%%%%%%%
\section{Rational Number Assignment for Retrograde Periodic Orbits }
\label{appendix:B}
%%%%%%%%%%%%%%%%%%%%%%%%%%%%%%%%%%%%%%%%%%%
Reference \cite{Levin:2008mq} provides a detailed theoretical analysis of the relationship between prograde periodic orbits and rational numbers. The accumulated azimuth in a radial period is given by
\bea
\Delta \varphi_{r+} = 2 \pi ( 1 + w + \frac{v}{z}).
\label{eq:angle}
\eea
We then associate a rational number with a prograde periodic orbit, where the two corresponding orbital frequencies are given by
\bea
w_r = \frac{2 \pi}{T_r}, \qquad w_\varphi = \frac{1}{T_r} \int_0^{T_r} \frac{d \varphi}{d t}\,\mathrm{d}t = \frac{\Delta \varphi_{r+}}{T_r},
\eea
where $T_r$ is the time spent on a radial cycle, which is a coordinate time. $w_r$ and $w_\varphi$ represent radial frequency and angular frequency, respectively. Then we can get
\bea
\frac{w_\varphi}{w_r} = \frac{\Delta \varphi_{r+}}{2 \pi} = 1 + w + \frac{v}{z} = 1 + q,
\eea
But for a retrograde periodic orbit, the accumulated azimuth is a negative accumulation, so we should change \eqref{eq:angle} to 
\bea
\Delta \varphi_{r-} = -2 \pi ( 1 + w + \frac{v}{z}).
\eea
So the relationship between retrograde periodic orbits and rational numbers should satisfy the following function
\bea
\frac{w_\varphi}{w_r} = \frac{\Delta\varphi_{r-}}{2 \pi}  = -w -\frac{v}{z} - 1 = q_- - 1.
\eea

\bibliographystyle{utphys}
\bibliography{refs}

\end{document}